\documentclass[review]{elsarticle}
\usepackage{lineno,hyperref}
\usepackage{graphics} 
\usepackage{graphicx} 
\usepackage{epstopdf}
\usepackage{color}
\usepackage{scrtime}
\usepackage{amssymb}
\usepackage{amsmath}
\usepackage{amsfonts}
\usepackage{natbib}
\usepackage[T1]{fontenc}

\modulolinenumbers[5]

\journal{Physics Letters B}

\begin{document}

\begin{frontmatter}

\title{Probing the $\mathrm{Z}=6$ spin-orbit shell gap with (p,2p) quasi-free scattering reactions}

\author[2,3]{I.~Syndikus}
\author[29]{M.~Petri\corref{mycorrespondingauthor}}
\cortext[mycorrespondingauthor]{Corresponding author}
\ead{marina.petri@york.ac.uk}
\author[LBNL]{A.~O.~Macchiavelli}
\author[29]{S.~Paschalis}
\author[6]{C.~A.~Bertulani}
\author[2,3]{T.~Aumann}
\author[1]{H.~Alvarez-Pol}
\author[2]{L.~Atar}
\author[1,MSU]{S.~Beceiro-Novo}
\author[1]{J.~Benlliure}
\author[1]{J.~M.~Boillos}
\author[3]{K.~Boretzky}
\author[7]{M.~J.~G.~Borge}
\author[NSCL,MSU]{B.~A.~Brown}
\author[1]{M.~Caama\~{n}o}
\author[2,3]{C.~Caesar}
\author[8]{E.~Casarejos}
\author[9]{W.~Catford}
\author[4]{J.~Cederkall}
\author[11]{L.~V.~Chulkov}
\author[1]{D.~Cortina-Gil}
\author[12]{E.~Cravo}
\author[13]{R.~Crespo}
\author[3,14]{I.~Dillmann}
\author[1,15]{P.~D\'{i}az Fern\'{a}ndez}
\author[16]{Z.~Elekes}
\author[2]{J.~Enders}
\author[3]{F.~Farinon}
\author[18]{L.~M.~Fraile}
\author[24]{D.~Galaviz Redondo}
\author[3]{H.~Geissel}
\author[21]{R.~Gernh{\"a}user}
\author[4]{P.~Golubev}
\author[22]{K.~G{\"o}bel}
\author[3]{M.~Heil}
\author[23]{M.~Heine}
\author[15]{A.~Heinz}
\author[24]{A.~Henriques}
\author[2]{M.~Holl}
\author[15]{H.~T.~Johansson}
\author[15]{B.~Jonson}
\author[25]{N.~Kalantar-Nayestanaki}
\author[26]{R.~Kanungo}
\author[3]{A.~Kelic-Heil}
\author[2]{T.~Kr{\"o}ll}
\author[3]{N.~Kurz}
\author[22]{C.~Langer}
\author[21]{T.~Le Bleis}
\author[24]{J.~F.~D.~C.~Machado}
\author[2,3,28]{J.~Marganiec-Ga\l \k{a}zka}
\author[7]{E.~Nacher}
\author[15]{T.~Nilsson}
\author[3]{C.~Nociforo}
\author[2]{V.~Panin}
\author[7]{A.~Perea}
\author[3]{S.~B.~Pietri}
\author[3]{R.~Plag}
\author[22]{R.~Reifarth}
\author[NSCL]{A.~Revel}
\author[7]{G.~Ribeiro}
\author[25]{C.~Rigollet}
\author[2,3]{D.~M.~Rossi}
\author[3]{D.~Savran}
\author[2]{H.~Scheit}
\author[3]{H.~Simon}
\author[31]{O.~Sorlin}
\author[7]{O.~Tengblad}
\author[32]{Y.~Togano}
\author[31]{M.~Vandebrouck}
\author[2,11,34]{V.~Volkov}
\author[2,3]{F.~Wamers}
\author[19]{C.~Wheldon}
\author[9]{G.~L.~Wilson}
\author[3]{J.~S.~Winfield}
\author[3]{H.~Weick}
\author[17]{P.~Woods}
\author[5]{D.~Yakorev}
\author[15]{M.~Zhukov}
\author[33]{A.~Zilges}
\author[30]{K.~Zuber}
\author[]{for the R$^3$B Collaboration}

\address[2]{Institut f{\"u}r Kernphysik, Technische Universit{\"a}t Darmstadt, Darmstadt, Germany}
\address[3]{GSI Helmholtzzentrum f{\"u}r Schwerionenforschung, Darmstadt, Germany}
\address[29]{Department of Physics, University of York, York, United Kingdom}
\address[LBNL]{Nuclear Science Division, Lawrence Berkeley National Laboratory, Berkeley, USA}
\address[6]{Texas A\&M University-Commerce, Commerce, USA}
\address[1]{IGFAE, Universidade de Santiago de Compostela, Santiago de Compostela, Spain}
\address[MSU]{Department of Physics and Astronomy, Michigan State University, East Lansing, USA}
\address[7]{Instituto de Estructura de la Materia, CSIC, Madrid, Spain}
\address[NSCL]{National Superconducting Cyclotron Laboratory, Michigan State University, East Lansing, USA}
\address[4]{Department of Physics, Lund University, Lund, Sweden}
\address[5]{Helmholtz-Zentrum Dresden-Rossendorf, Institute of Radiation Physics, Dresden, Germany}
\address[8]{Universidad de Vigo, Vigo, Spain}
\address[9]{Department of Physics, University of Surrey, Surrey, United Kingdom}
\address[11]{National Research Centre ``Kurchatov Institute'', Moscow, Russia}
\address[34]{NRC ``Kurchatov Institute'' - ITEP, Moscow, Russia}
\address[12]{Departamento de F\'{i}sica, Instituto Superior T\'{e}cnico, Lisboa, Portugal}
\address[13]{Instituto Superior Tecnico, University of Lisbon, Lisboa, Portugal}
\address[14]{Justus-Liebig-Universit{\"a}t Gie\ss en, Gie\ss en, Germany}
\address[15]{Chalmers University of Technology, G{\"o}teborg, Sweden}
\address[16]{ATOMKI Debrecen, Bem t\'{e}r 18/c, Debrecen, Hungary}
\address[17]{Department of Physics, University of Edinburgh, Edinburgh, United Kingdom}
\address[18]{Grupo de F\'{i}sica Nuclear \& IPARCOS, Universidad Complutense de Madrid, Madrid, Spain}
\address[21]{Physik Department E12, Technische Universit{\"a}t M{\"u}nchen, Garching, Germany}
\address[22]{Johann Wolfgang Goethe-Universit{\"a}t Frankfurt, Frankfurt am Main, Germany}
\address[23]{IPHC Strasbourg, France}
\address[24]{Nuclear Physics Center, University of Lisbon, Lisboa, Portugal}
\address[25]{KVI-CART, University of Groningen, Groningen, Netherlands}
\address[26]{Saint Mary's University, Halifax, Nova Scotia, Canada}
\address[27]{Science and Technology Facilities Council - Daresbury Laboratory, Warrington, United Kingdom}
\address[28]{Extreme Matter Institute, GSI Helmholtzzentrum f{\"u}r Schwerionenforschung, Darmstadt, Germany}
\address[30]{Technische Universit{\"a}t Dresden, Institut f{\"u}r Kern- und Teilchenphysik, Dresden, Germany}
\address[31]{GANIL, Bd Henri Becquerel, Caen, France}
\address[32]{RIKEN, Nishina Center for Accelerator-Based Science, Wako, Saitama, Japan}
\address[33]{Institut f{\"u}r Kernphysik, Universit{\"a}t zu K{\"o}ln, K{\"o}ln, Germany}
\address[19]{Department of Physics, University of Birmingham, Birmingham, United Kingdom}

\begin{abstract}
The evolution of the traditional nuclear magic numbers away from the valley of stability is an active field of research. Experimental efforts focus on providing key spectroscopic information that will shed light into the structure of exotic nuclei and understanding the driving mechanism behind the shell evolution. In this work, we investigate the $\mathrm{Z}=6$ spin-orbit shell gap towards the neutron dripline. To do so, we employed $\rm ^AN$(p,2p)$\rm ^{A-1}C$
quasi-free scattering reactions to measure the proton component of the $2^+_1$ state of $^{16,18,20}$C.
The experimental findings support the notion of a moderate reduction of the proton $1p_{1/2} - 1p_{3/2}$ spin-orbit splitting, at variance to recent claims for a prevalent $\mathrm{Z}=6$ magic number towards the neutron dripline. 
\end{abstract}


\end{frontmatter}

\section{Introduction \label{Introduction}}

The emergence of nuclear magic numbers within a shell-model description of atomic nuclei has 
been the paradigm of our understanding of nuclear structure.
The theoretical interpretation of these special numbers followed from the inclusion of a strong spin-orbit force in the nuclear mean-field potential \cite{PhysRev.74.235, PhysRev.75.1969, PhysRev.75.1766.2}. However, due to the isospin dependence of the nucleon-nucleon (NN) residual interaction, the magic numbers that emerge near the stability line are not necessarily the same for exotic nuclei, which
have large neutron-proton asymmetry, see e.g. Ref.~\cite{SORLIN2008602} and references therein.

The first spin-orbit shell gap originates from the splitting of the $1p_{1/2} -1p_{3/2}$ orbits and the $\mathrm{Z}=6$ carbon isotopes provide an excellent case to study the changes in the proton spin-orbit splitting from stability to the dripline.
This case is particularly interesting since it is anticipated that a quenching of the splitting will occur due to effects of the tensor and two-body spin-orbit forces acting on the $1p$ protons when neutrons are added in the $d_{5/2}$ and $s_{1/2}$ orbits.
Indeed, in our earlier work \cite{Macchiavelli2014} we have attributed the observed  increase in the $B(E2; 2^{+}_{1} \rightarrow 0^{+}_{1})$ values from $^{16}$C to $^{20}$C, as a manifestation of increased in-shell proton excitations ($p^1_{1/2}p^{-1}_{3/2}$), due to a weakening of  the $1p_{1/2} - 1p_{3/2}$ spin-orbit splitting at $\mathrm{Z}=6$ towards the dripline. 
It is worth pointing out a similar effect has been discussed in Ref.~\cite{FEDERMAN1984269} for the case of the $2p_{1/2} - 2p_{3/2}$ splitting in the yttrium isotopes. 

Following on from Refs.~\cite{Macchiavelli2014, Petri2011, Petri2012}, we report in this Letter the results of an experiment designed to study the proton component of the $2^+_1$ state in $^{16,18,20}$C using Quasi-Free Scattering (QFS) (p,2p) reactions on $^{17,19,21}$N.
As described in more detail in Section~\ref{Discussion}, the proton component ($p^1_{1/2}p^{-1}_{3/2}$) in the carbon isotopic chain can be uniquely accessed through (p,2p) reactions;
the  population of the $2^+$ state is expected to proceed only through this component since the $\pi 1p_{1/2}$ coupled to the $2^+$ state in the carbon isotopes cannot contribute to the $1/2^-$ ground state of the nitrogen isotopes.
Our results show an increase in the proton component, and signals a quenching of the $\mathrm{Z}=6$ $1p_{1/2} - 1p_{3/2}$ gap towards the dripline, casting doubt on the strong statement given by the authors of Ref.~\cite{Tran2018}, who conclude that the $\mathrm{Z}=6$ spin-orbit originated magic number is prevalent up to $^{20}$C.

\section{Experimental Details \label{Experimental Details}}

The experiment was performed at GSI Helmholtzzentrum f\"ur Schwerionenforschung using the R$^3$B/LAND setup \cite{R3BSetup}.
A primary beam of $^{40}$Ar was used to bombard a 4\, g/cm$^2$ Be production target with an energy of 490\,MeV/nucleon.
The products were selected by the fragment separator FRS \cite{Geissel1992} according to their $B\rho$ and delivered to the R$^3$B/LAND setup.

\begin{figure*}[tbp]
 \centering
    \includegraphics[angle=0, width=0.49\textwidth, trim=0 20 0 0, clip=true]{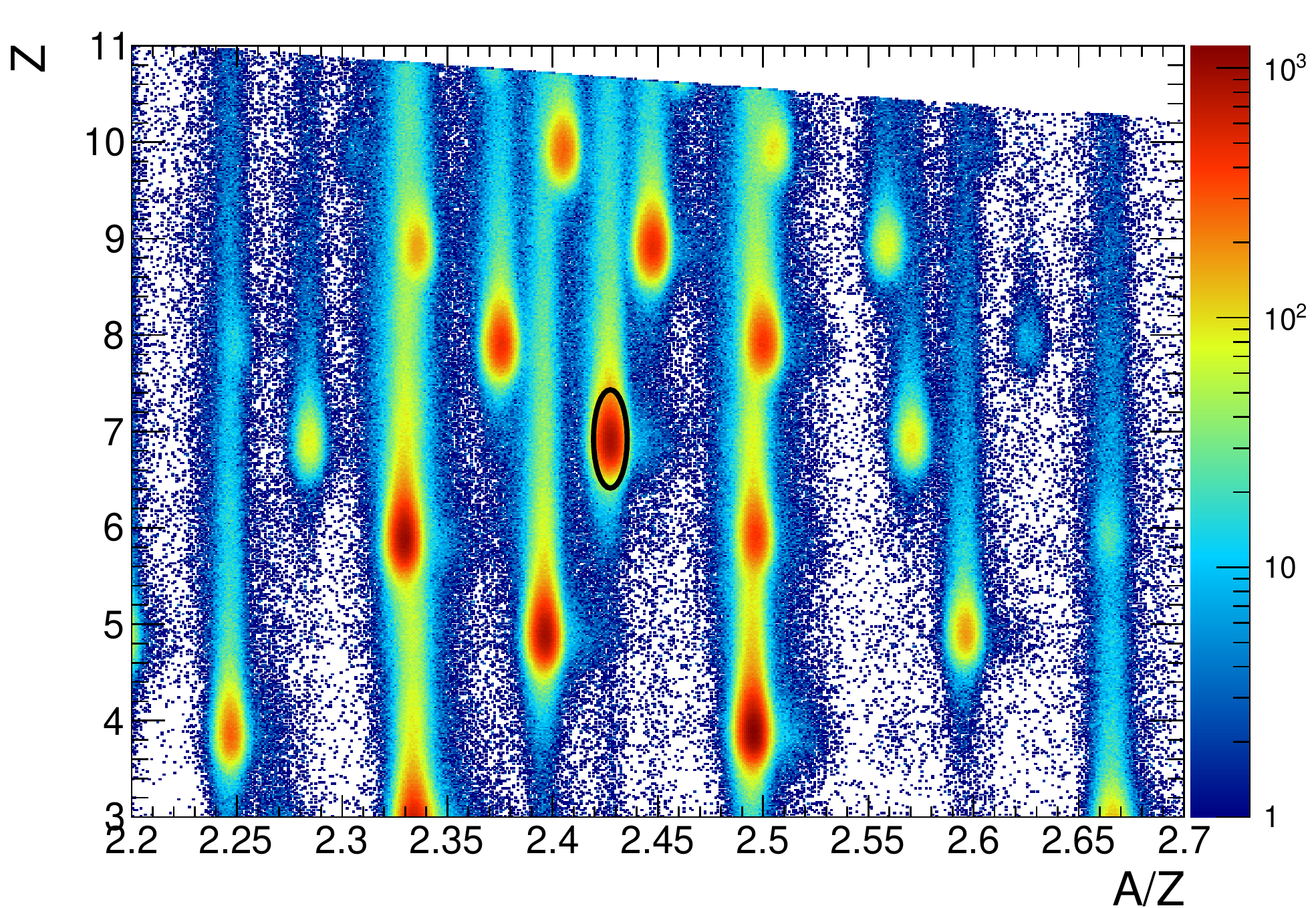}
    \includegraphics[angle=0, width=0.49\textwidth, trim=0 20 0 0, clip=true]{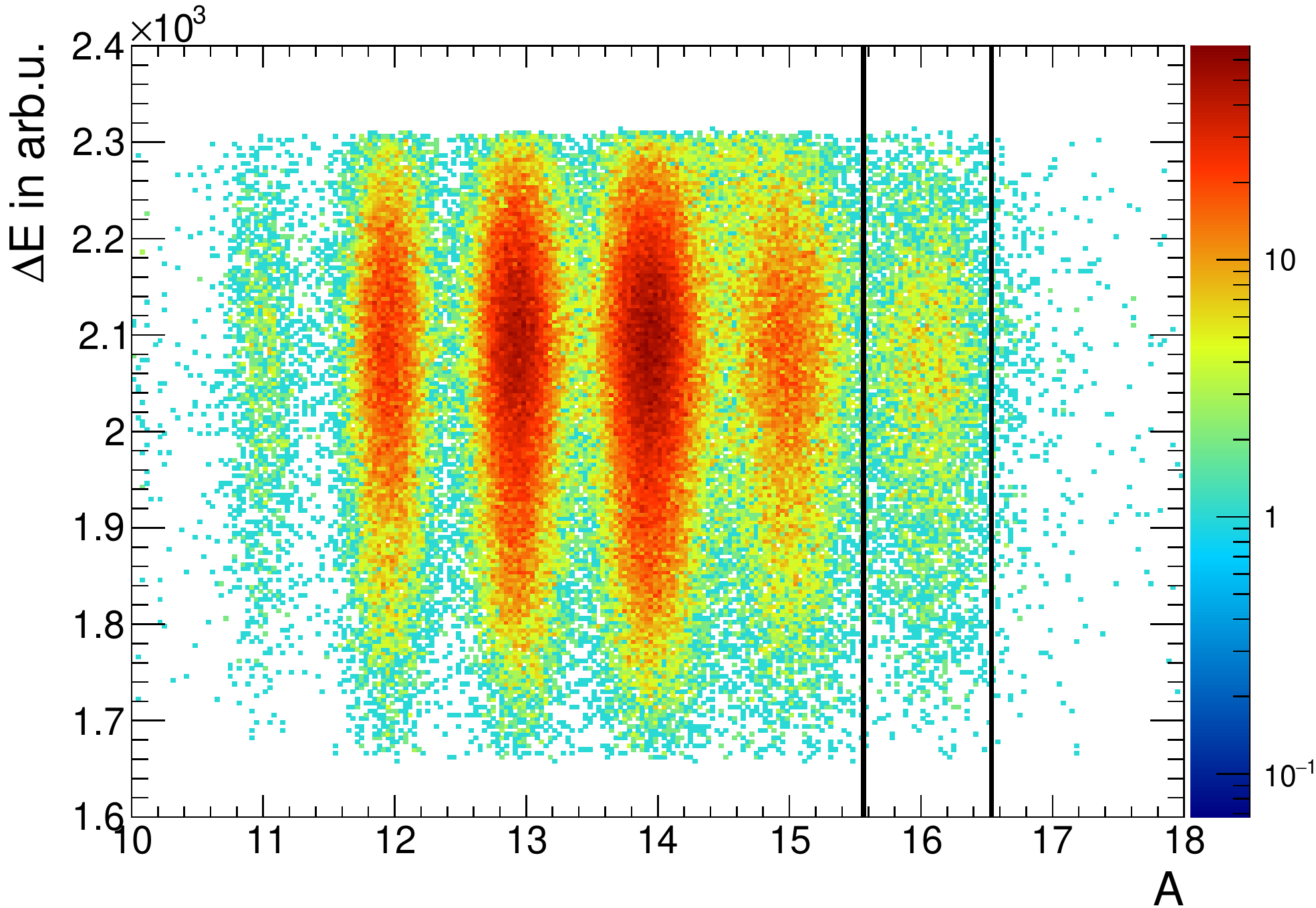}
    \includegraphics[angle=0, width=0.49\textwidth, trim=0 20 0 0, clip=true]{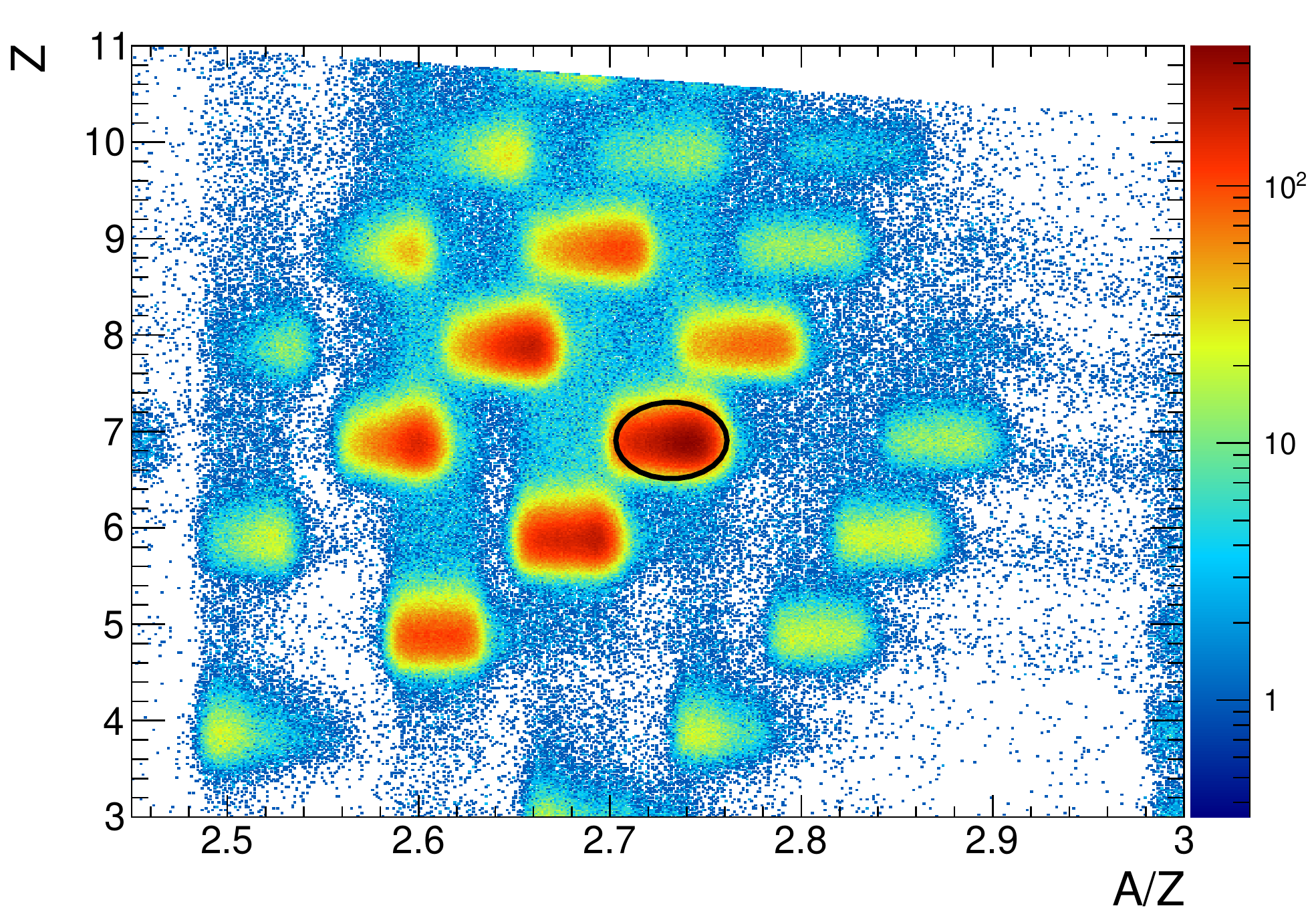}
    \includegraphics[angle=0, width=0.49\textwidth, trim=0 20 0 0, clip=true]{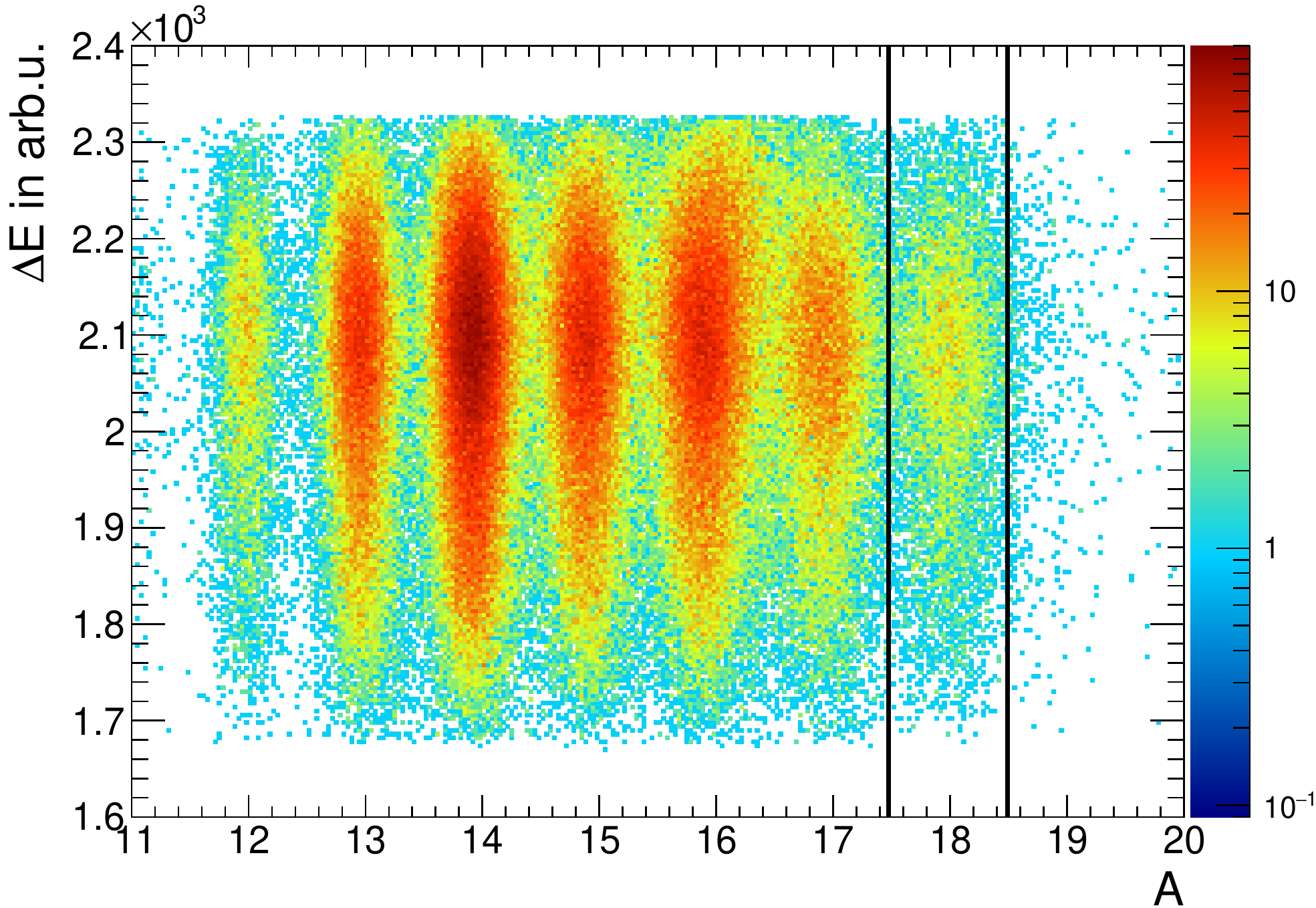}
   \includegraphics[angle=0, width=0.49\textwidth, trim=0 0 0 0, clip=true]{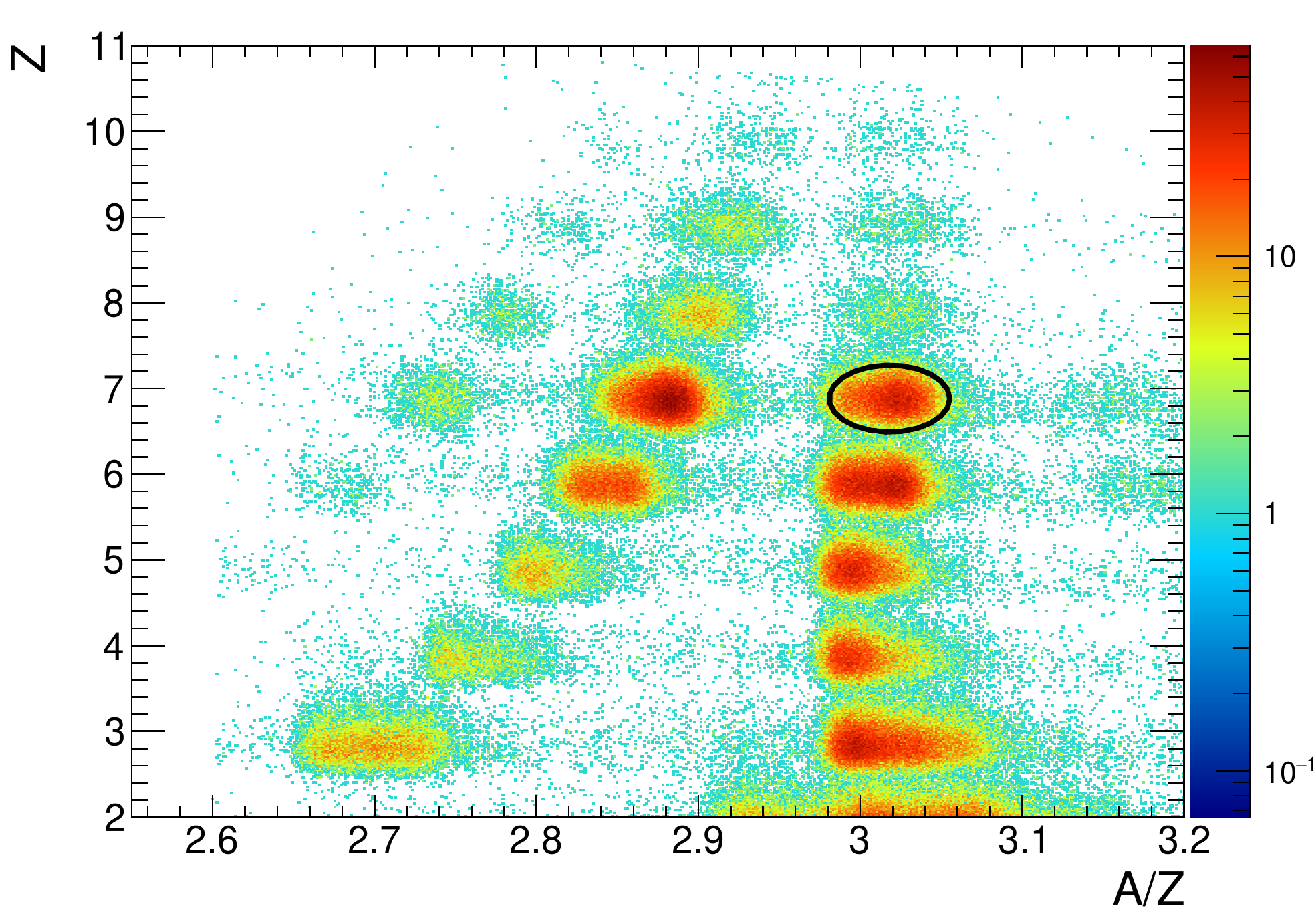}
  \includegraphics[angle=0, width=0.49\textwidth, trim=0 0 0 0, clip=true]{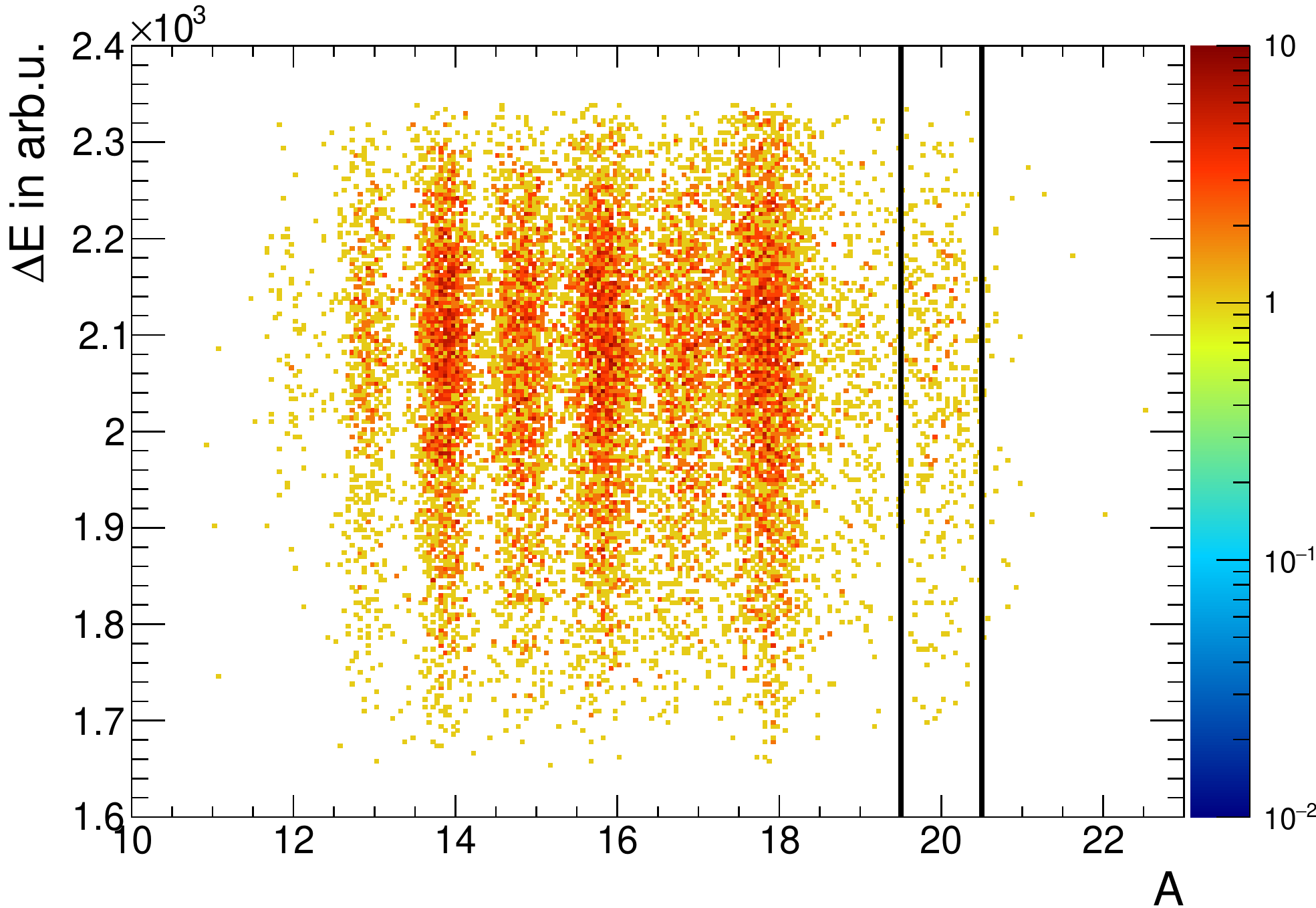}
    \caption{(Color online) Incoming (all $\rm Z$) (left) and outgoing ($\rm Z=6$) (right) PID plots for the reaction $^{17}$N(p,2p)$^{16}$C (first row), $^{19}$N(p,2p)$^{18}$C (second row), and $^{21}$N(p,2p)$^{20}$C (third row) using a CH$_2$ reaction target. The gates on the isotopes of interest are marked with an ellipsoid (incoming PID plots) and two straight lines (outgoing PID plots). \label{Fig_PID}}
\end{figure*}

The R$^3$B/LAND setup
enables a kinematically complete measurement of QFS (p,2p) reactions in inverse kinematics.
The incoming isotopes are fully stripped and identified on an event-by-event basis via energy-loss, time-of-flight and position measurements. 
The outgoing heavy charged fragments are bent by the dipole magnet ALADIN towards the fragment arm
and their $\mathrm{A/Z}$ ratio is determined by their trajectory through the magnetic field.
As an example the particle-identification (PID) plots for all incoming isotopes and outgoing (only $\rm Z=6$) fragments together with the applied software gates are shown in Figure~\ref{Fig_PID}.
The protons from the QFS reactions and $\gamma$ rays from the decay of excited states are detected by the Crystal Ball (XB) detector array \cite{Metag1982} surrounding the target area.
More details on the experimental setup can be found in Refs.~\cite{DiazFernandez2013, Holl2014,Atar2018,DiazFernandez2018,HOLL2019682}.

\begin{figure}[tbp]
 \centering
    \includegraphics[angle=0, width=0.49\textwidth, trim=0 0 0 0, clip=true]{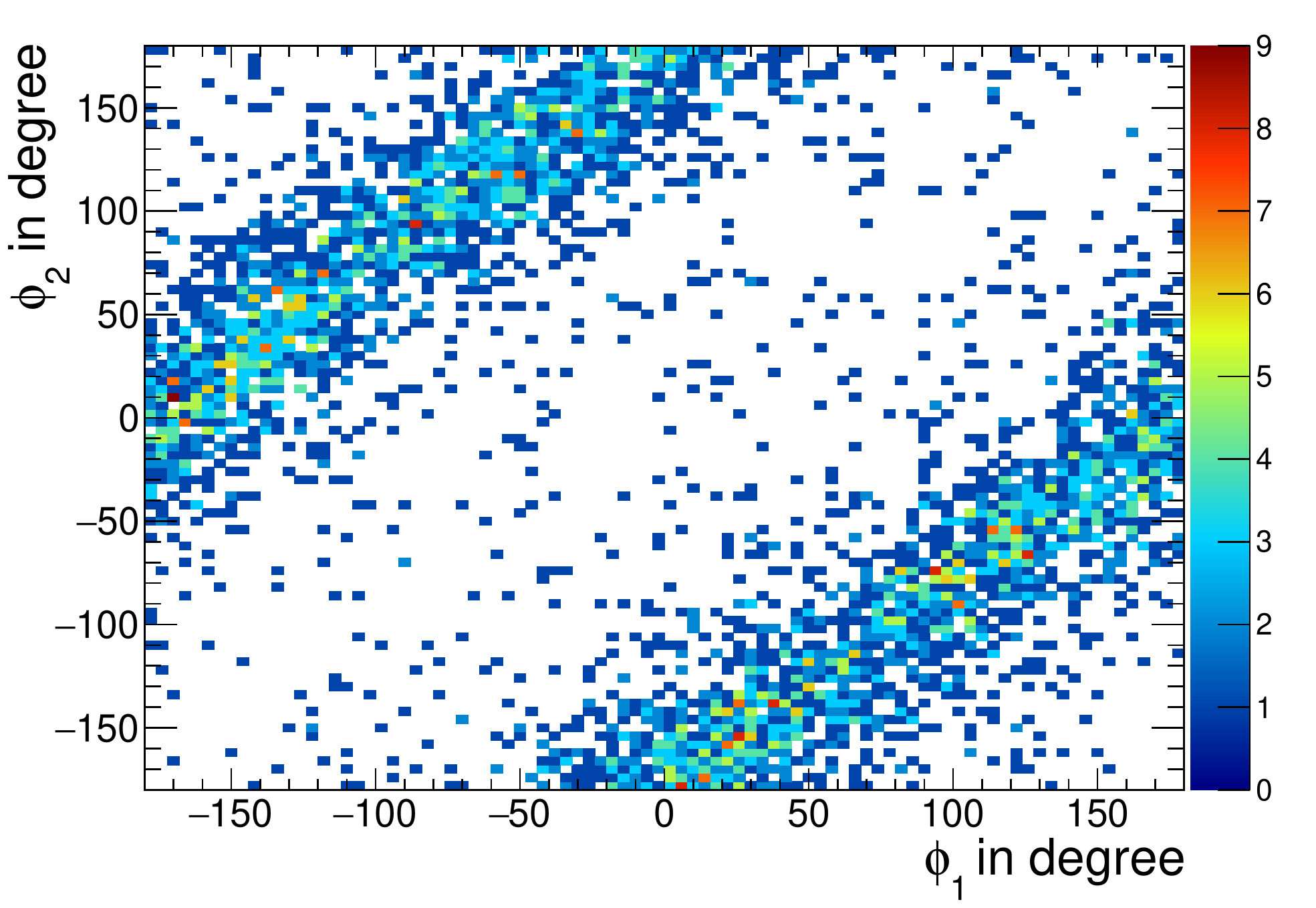}\\
    \includegraphics[angle=0, width=0.49\textwidth, trim=0 0 0 0, clip=true]{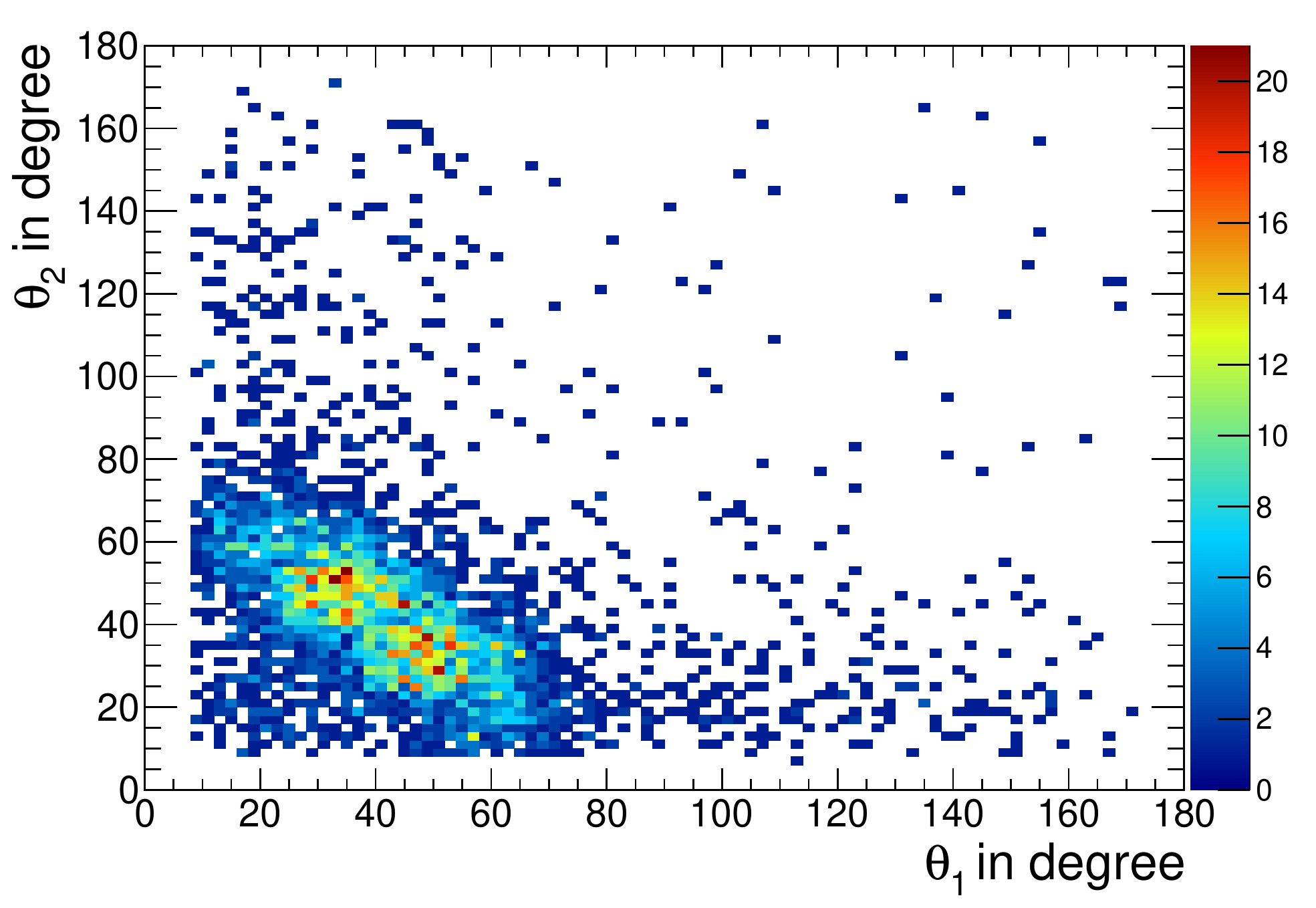}
    \caption{(Color online) Angular correlations of the two quasi-free scattered protons on the CH$_2$ target for the $^{19}$N(p,2p)$^{18}$C reaction. In the laboratory frame, the azimuthal angles $\phi$ (top) show the expected back-to-back nearly co-planar scattering, while the polar angles $\theta$ (bottom) indicate that protons scatter with $\Delta\theta \approx\ 80^\circ$, as expected for the QFS reactions at these energies. \label{Fig_Protons}}
\end{figure}

For this work, we study (p,2p) reactions from $^{17,19,21}$N to $^{16,18,20}$C, respectively.
To emulate reactions on a pure proton target, data were taken for both CH$_2$ and C targets.
The contribution from the protons in the CH$_2$ target is then reconstructed by subtracting the contribution from the C target.
In addition, a measurement without target (empty target) was performed to estimate the contribution from the in-beam detectors to the reaction of interest. The target properties as well as the energy of the incoming isotopes are listed in Table~\ref{Table_Properties}.
With a gate on the incoming $\rm^A$N and the outgoing $\rm^{A-1}$C isotopes, the reactions of interest are selected.
In addition to the fragment selection, the identification of the (p,2p) QFS reactions is performed by gating on the two outgoing protons with XB.
A next-neighbour addback around the proton hit is applied in the data analysis to remove contamination from the proton hits into the $\gamma$-ray spectra.
The distribution of the protons in the laboratory frame, shown in Figure~\ref{Fig_Protons}, 
manifests the characteristics of the QFS reactions.

For the identification of the bound excited states in the carbon isotopes, the emitted $\gamma$ rays are detected with XB.
To reduce the Compton background, the crystals in which some energy was deposited are sorted into clusters.
A cluster is built by taking the highest energy entry and adding the energy of all neighbouring crystals (next-neighbour addback).
Each cluster then corresponds to a single $\gamma$ ray.
Afterwards, its energy is Doppler corrected taking into account the position of the cluster center.

\begin{table}[tbp]
\centering
\caption{\label{Table_Properties} Incoming beam and target properties.}
\begin{tabular}{c c c || c c}\hline\hline
Beam & Energy [MeV/nucleon] & Total \# of ions  & Target & Thickness [g/cm$^2$]\\\hline\hline
$^{17}$N & 438 & $3.533\times10^7$  & CH$_2$ & 0.458 \\
& & $1.131\times10^7$ &  C & 0.558 \\
$^{19}$N & 430 & $2.534\times10^7$  & CH$_2$ & 0.922 \\
& & $1.034\times10^7$ &  C & 0.935 \\
$^{21}$N & 422 & $1.693\times10^6$ &   CH$_2$&  0.922 \\
& & $5.127\times10^5$ &  C & 0.935 \\\hline\hline
\end{tabular} 
\end{table}

\section{Results \label{Results}}

To obtain inclusive and exclusive cross sections in the $^{17,19,21}$N(p,2p)$^{16,18,20}$C reactions,
one needs to know the $\gamma$-ray and proton efficiencies of XB.
These are determined with simulations, using the R3BRoot framework \cite{R3BRoot} and the Geant4 v10.2.1 transport engine \cite{Agostinelli2003}.
The $\gamma$-ray photopeak efficiency of XB was simulated and compared with available source measurements ($^{60}$Co).
In addition, detailed studies have been performed \cite{DiazFernandez2018} to check the validity of these simulations with respect to the detection of both neutrons and protons. 

For the production of the particles of interest in the simulation, namely the outgoing fragment and the two protons from the QFS reaction, an external event generator is used.
The event generator is based on a pure kinematical description of the reaction process assuming an isotropic center-of-mass collision \cite{Chulkov2005}.
The simulation of 100,000 events leads to a two-proton efficiency of 
$\epsilon_\mathrm{2p} = 57.4(3)\%, 56.2(3)\%$ and $55.4(3)\%$ for the $^{17}$N(p,2p)$^{16}$C, $^{19}$N(p,2p)$^{18}$C and $^{21}$N(p,2p)$^{20}$C reactions, respectively. The decrease in $\epsilon_\mathrm{2p}$ is due to the decrease in the energy of the incoming nitrogen isotopes from $^{17}$N to $^{21}$N.

In addition, the simulations are used to extract the contributions from the bound excited states to the $\gamma$-ray spectra.
The same event generator as for the proton efficiency is used, with the relevant $\gamma$ rays from the de-excitation added to the fragment and the two protons.
Each decay channel is simulated separately to take side-feeding of the lower-lying state into account.

\begin{figure*}[tbp]
 \centering
    \includegraphics[angle=0, width=\textwidth, trim=0 20 0 0, clip=true]{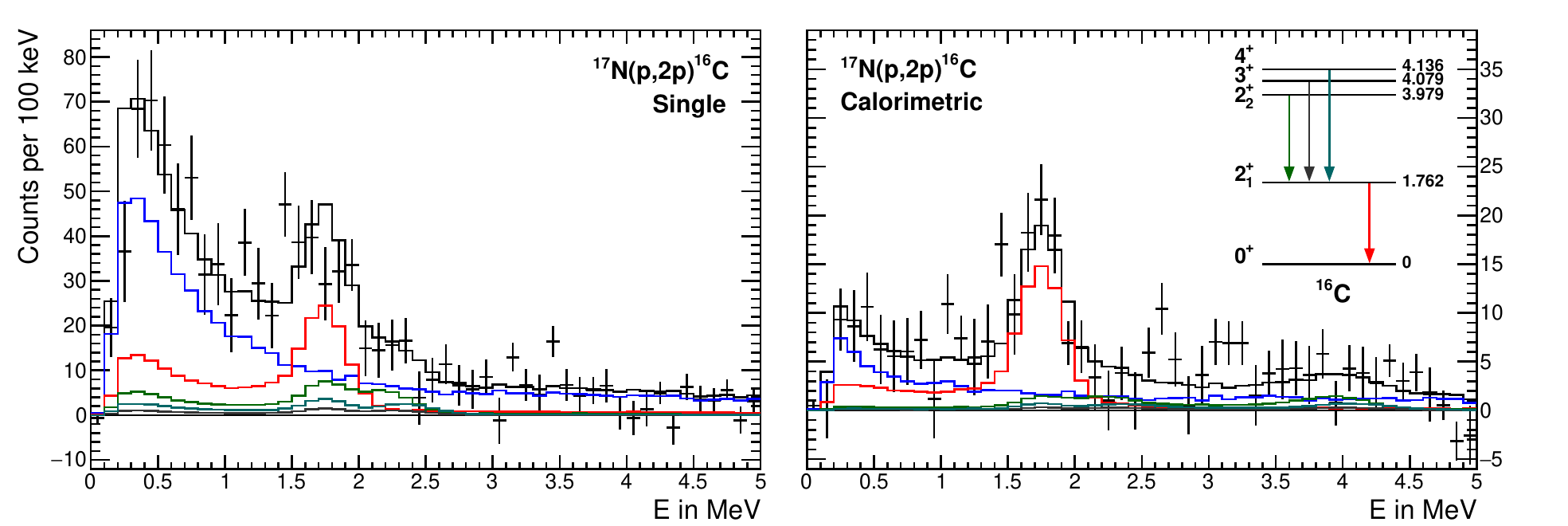}
    \includegraphics[angle=0, width=\textwidth, trim=0 20 0 0, clip=true]{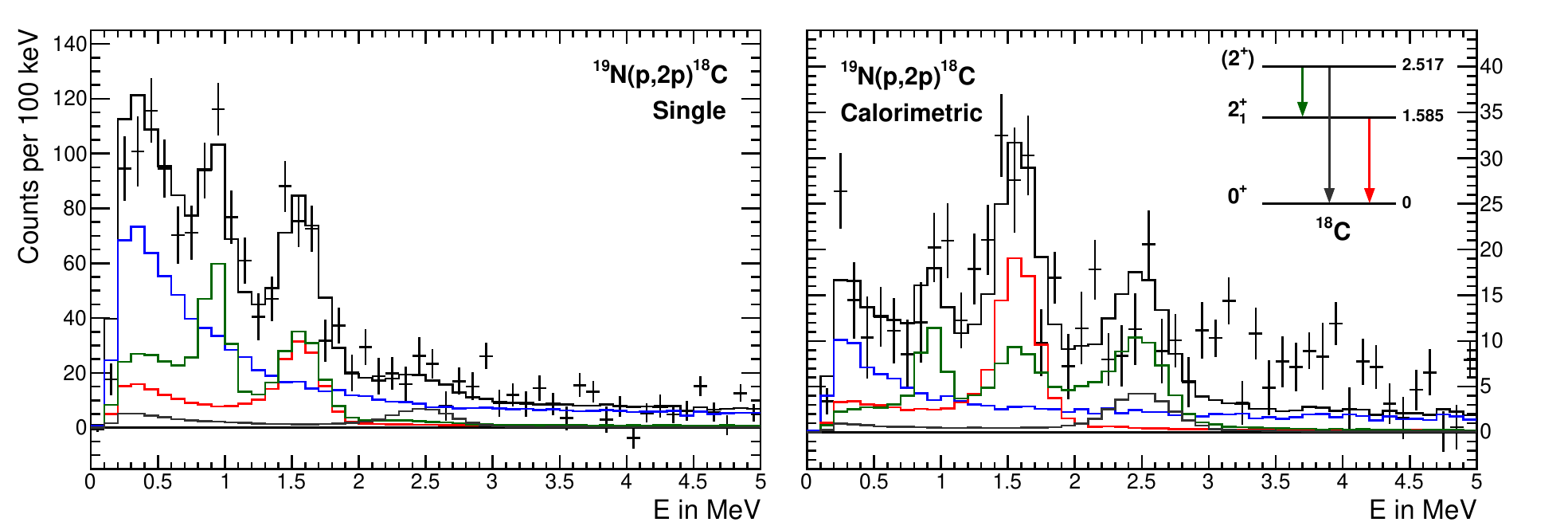}
    \includegraphics[angle=0, width=\textwidth, trim=0 0 0 0, clip=true]{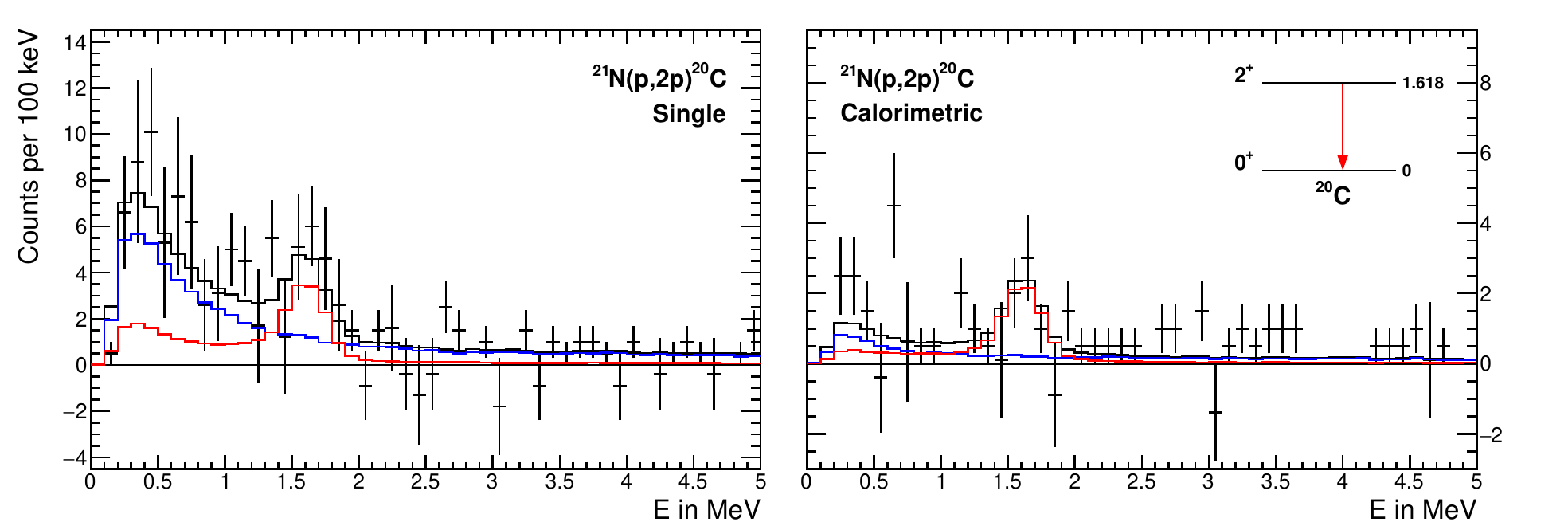}
    \caption{(Color online) Single (left) and calorimetric (right) $\gamma$-ray spectra for the reactions $^{17}$N(p,2p)$^{16}$C (first row), $^{19}$N(p,2p)$^{18}$C (second row), and $^{21}$N(p,2p)$^{20}$C (third row) on the reconstructed H target. The experimental spectra are shown with black crosses. In all cases, the simulation is depicted for the background induced by the protons (blue) and the $\gamma$ rays de-exciting the first $2^{+}$ state (red). In addition, for $^{17}$N(p,2p)$^{16}$C the
    de-excitation of the higher-lying $2_2^{+}$ (green), $3^{+}$ (grey), and $4^{+}$ (turquoise) states to the first $2^{+}$ state is also considered in the simulations. For $^{19}$N(p,2p)$^{18}$C the direct decay of the $2^{+}_2$ state to the ground state (grey) and its decay via a cascade (green) are shown. The sum of all simulations is shown in black. The insets show the level scheme and $\gamma$-ray transitions (in color) as considered in the simulations. \label{Fig_Spectra} }
\end{figure*}

For the $^{17}$N(p,2p)$^{16}$C reaction, the level scheme of $^{16}$C shown in Figure~\ref{Fig_Spectra} is considered, as observed in \cite{Petri2012} following a proton-removal reaction. A small contribution from the direct decay of the second 2$^+$ could remain undetected, however, this direct decay has been constrained to a branching ratio of less than 8.8\% in \cite{Petri2012}, and therefore the error induced by omitting this is negligible.

For the $^{19}$N(p,2p)$^{18}$C reaction, the level scheme of $^{18}$C as shown in Figure~\ref{Fig_Spectra} is considered. 
Since the direct decay of the $2_2^+$ state to $0^+$ ground state is simulated separately from the decay via the $2_1^+$ state, the branching ratio can be extracted by comparing the contributions from the two decay paths from the fit shown in Figure~\ref{Fig_Spectra}.
The branching ratio of the direct decay is $22(8)\,\%$, while the one of the cascade is $78(8)\,\%$.
This is in good agreement with the value determined in one-proton removal reactions \cite{Voss2012}.
The calorimetric spectrum shows a possible peak at 4~MeV, which would be in line with a level observed by Stanoiu et al. \cite{Stanoiu2008} in a multi-fragmentation reaction and Kondo et al. \cite{Kondo2009} following a neutron-removal reaction (with a large spectroscopic factor, suggesting that this state is dominated by a neutron configuration). However, in this work, we are looking into the very selective proton-removal channel, and therefore not all states known for $^{18}$C are expected to be populated. We followed the level-scheme of Ref.~\cite{Voss2012}, where the 4~MeV state is not populated following a proton-removal reaction, and we did not include this state in our analysis, treating this as background. If we do include this state in our analysis, the proton amplitude decreases only slightly (from 7\% to 5\%) leaving the conclusions of this work unaffected.

For the $^{21}$N(p,2p)$^{20}$C reaction, the level scheme of $^{20}$C as shown in Figure~\ref{Fig_Spectra} is considered. 
For $^{20}$C only one bound $2_1^+$ excited state at 1.6~MeV has been considered, consistent with 
the $\gamma$-ray spectrum observed in Ref.~\cite{Petri2011}.

A combination of the simulated spectra of all bound excited states are then fitted to the experimental ones.
The single and calorimetric spectra are fitted simultaneously using Neyman's $\chi^2$ estimator \cite{Baker1984}.
The single spectra are filled with the energy entries of all clusters separately.
For the calorimetric spectra, the energies of all clusters for a given event are summed.
Since the two types of spectra are sensitive to the different decays -- direct or via a cascade -- in different ways, the ratio of the excited states can be determined more accurately by fitting both spectra simultaneously.
The results are shown in Figure~\ref{Fig_Spectra}.

The inclusive cross section can be calculated taking into account the proton detection efficiency from the simulation.
The results are given in Table~\ref{Table_CrossSections} for the three reactions of interest.

The number of events in a certain excited state is determined from a fit of the simulated to the experimental $\gamma$-ray spectra, as depicted in Figure~\ref{Fig_Spectra}.
The resulting exclusive cross sections for the $0^+$ ground state and the $2_1^+$ excited state are listed in Table~\ref{Table_CrossSections}.
The cross section of the ground state is calculated by subtracting all excited states from the inclusive cross section.
The exclusive cross sections for the higher-lying states are not listed but can be reconstructed from the difference of the inclusive and exclusive cross sections listed in Table~\ref{Table_CrossSections}.

The inclusive cross section for the $^{21}$N(p,2p)$^{20}$C reaction has been published by our collaboration in an independent analysis \cite{DiazFernandez2018} using the same data set. 
The two cross sections ($\sigma_\mathrm{kin} = 2.27(38)$ from \cite{DiazFernandez2018} and $\sigma = 2.65(34)$ from this work) are consistent within their statistical uncertainty (of $\sim$~15\%) and the systematic uncertainty (of $\sim$~6\%) induced by the choice of thresholds in the addback procedure as discussed in detail in Ref.~\cite{HOLL2019682}. 
Moreover, for this work, the relative exclusive cross sections along the C isotopic chain is of importance, therefore systematic deviations in their absolute value, due to e.g. different thresholds, do not impact the conclusion of this work. 

\section{Discussion \label{Discussion}}

The experimental cross sections in Table~\ref{Table_CrossSections} are compared to theoretical predictions based on the eikonal approximation \cite{Aumann2013}, well suited for (p,2p) QFS reactions at the energies considered here.
The optical potential is given by the $t\rho\rho$ approach.
The density of the nuclei are determined with Hartree-Fock calculations using the SLy5 interaction \cite{Chabanat1998} and isovector surface pairing.
The wave function of the proton is calculated using a phenomenological potential consisting of a Wood-Saxon central potential, a spin-orbit term, and the Coulomb potential.
The binding energy of the proton in the $p_{1/2}$ orbit is given by the proton separation energy $S_\mathrm{p}$, while the binding of the proton in the $p_{3/2}$ orbit is estimated from the proton single-particle levels in $^{15}$N.
The theoretical exclusive single-particle cross sections for QFS on the $p_{1/2}$ and $p_{3/2}$ protons are given in Table~\ref{Table_CrossSections}. When combined with the experimentally derived exclusive cross sections, we can calculate the experimental spectroscopic factors, $C^2S$, for the $0_1^+$ and $2_1^+$ states from

\begin{align*}
\sigma_\mathrm{exp}=C^2S\sigma_\mathrm{theo}.
\end{align*}

\begin{table*}[tbp]
\centering
\caption{\label{Table_CrossSections} Experimental and theoretical cross sections and spectroscopic factors. The theoretical cross sections were calculated using the formalism described in Ref.~\cite{Aumann2013} for an occupation number of 1 for both the $1p_{1/2}$ and the $1p_{3/2}$ orbit (see text for details).} 
\begin{tabular}{c c c c c c c c}\hline\hline
  & State & Orbital & $\sigma_\mathrm{exp}$[mb] & $\sigma_\mathrm{theo}$[mb] & $C^2S_{exp}$ & $\beta^2$ [\%] \\ \hline\hline
  $^{17}$N(p,2p)$^{16}$C & inclusive &        & 3.82(19) &       &         &        \\    
                         & $0^+$ & $1p_{1/2}$ & 2.83(20) & 6.171 & 0.46(3) &          \\
                         & $2^+$ & $1p_{3/2}$ & 0.68(9)  & 5.929 & 0.11(2) & 10.0(15) \\ \hline
  $^{19}$N(p,2p)$^{18}$C & inclusive &        & 3.66(14) &       &         &        \\    
                         & $0^+$ & $1p_{1/2}$ & 2.53(15) & 5.267 & 0.48(3) &          \\
                         & $2^+$ & $1p_{3/2}$ & 0.45(7)  & 5.193 & 0.09(1) & 7.2(12)  \\ \hline
  $^{21}$N(p,2p)$^{20}$C & inclusive &        & 2.65(34) &       &         &        \\  
                         & $0^+$ & $1p_{1/2}$ & 1.87(38) & 4.554 & 0.41(8) &           \\
                         & $2^+$ & $1p_{3/2}$ & 0.78(17) & 4.458 & 0.17(4) & 17.0(51)  \\\hline\hline
\end{tabular}
\end{table*}

\noindent To determine the proton contribution to the $2_1^+$ state wave function from our cross section measurements, we consider
a simple shell-model picture for the $0_1^+$ and $2_1^+$ state as discussed in Refs.~\cite{Macchiavelli2014,Petri2012}. We assume that the $0^+$ ground state of the carbon isotopes can be described as
\begin{align*}
 |0_1^+;\, \rm^{A-1}C\rangle \approx  & \ |\nu\, (sd)^n;J=0\rangle \otimes |\pi\, (1p_{3/2})^{4};J=0\rangle,
\end{align*}
with $n=2,4,6$ for $^{16}$C, $^{18}$C, and $^{20}$C, respectively, with the valence neutrons occupying a quasi-degenerate $sd$ shell \cite{Stanoiu2008}.
Additionally, the $2^+_1$ excited state can be described as
\begin{align*}
 |2^+_1;\, \rm^{A-1}C\rangle  \approx &\ \alpha |\nu\, (sd)^n;J=2\rangle \otimes |\pi\, (p_{3/2})^{4};J=0\rangle \\ 
 & + \beta |\nu\, (sd)^n;J=0\rangle \otimes |\pi\, (p_{3/2})^{3}(p_{1/2})^{1};J=2\rangle,
\end{align*}
where $\alpha$ and $\beta$ denote the amount of pure neutron and pure proton excitation contributing to the state, respectively.
Naturally, within this simple scheme the ground state of the neutron-rich nitrogen isotopes is
\begin{align*}
 |1/2^-;\, \rm^AN\rangle  \approx & |\nu\, (sd)^n;J=0\rangle \\
 & \otimes |\pi\, (p_{3/2})^{4}(p_{1/2})^{1};J=1/2\rangle.
\end{align*}
This is supported by experimental data for $^{15,17}$N. In particular, spectroscopic factors for the $^{18}$O(d,$^3$He)$^{17}$N reaction~\cite{MAIRLE197797} confirmed that $1/2^-$ ground state contains the full 
$1p_{1/2}$ strength.  In addition, the magnetic moment of the ground state in $^{15}$N~\cite{RAGHAVAN1989}, -0.28 $\mu_N$, is very close to the Schmidt limit for the $\pi p_{1/2}$ orbit. That of $^{17}$N~\cite{UENO96}, -0.35 $\mu_N$, can be explained by a small component of $\approx$ 2\%
for the $|\nu\, (sd)^n;J=2^+\rangle \otimes |\pi\, (p_{3/2})^{3}(p_{1/2})^{2};J=3/2^-\rangle$ configuration in the ground state, because its contribution to the magnetic moment is large, $\approx$-5.2$\mu_N$.

Thus, and since the coupling of the $\pi 1p_{1/2}$ with the $2^+_1$ state of  the core cannot contribute to the ground state,
the spectroscopic factor for the removal of a proton in the $1p_{1/2}$ orbit populating the $0_1^+$ ground state of $\rm ^{A-1}C$
is expected to be 1.
The removal of a $1p_{3/2}$ proton from the ground state will populate the $2^+_1$ state only via the proton component and directly probes
the proton amplitude of the state, $\beta$.

With four protons in the $1p_{3/2}$ orbit, the ratio of the spectroscopic factors in our simple picture is proportional to the proton amplitude (see Ref.~\cite{Petri2012}):
\begin{align}
 \frac{\sigma_\mathrm{exp}(2^+_1)}{\sigma_\mathrm{exp}(0^+_1)} \times \frac{\sigma_\mathrm{theo}(p_{1/2})}{\sigma_\mathrm{theo}(p_{3/2})} = \frac{C^2S(2^+_1)}{C^2S(0_1^+)}  = \beta^2 \times \frac{5}{2}.
 \label{Equ_ProtonAmplitude}
\end{align}
In looking at the ratios,  quenching factors (see e.g. \cite{Atar2018,PASCHALIS2020135110,Tostevin2014,doi:10.1142/S0217751X09045625,PhysRevLett.103.202502, PhysRevLett.107.032501}) largely cancel out and can be compared directly to shell-model expectations.
Taking into account the experimental and  theoretical exclusive cross sections in Table~\ref{Table_CrossSections} and Eq.~\ref{Equ_ProtonAmplitude}, the corresponding proton amplitudes of the $2^+_1$ states in  $^{16,18,20}$C, $\beta^2$, can be calculated and are listed in Table~\ref{Table_CrossSections}.

\begin{figure}[tbp]
 \centering
 \includegraphics[angle=0, width=0.7\textwidth, trim=0 0 0 0,clip=true]{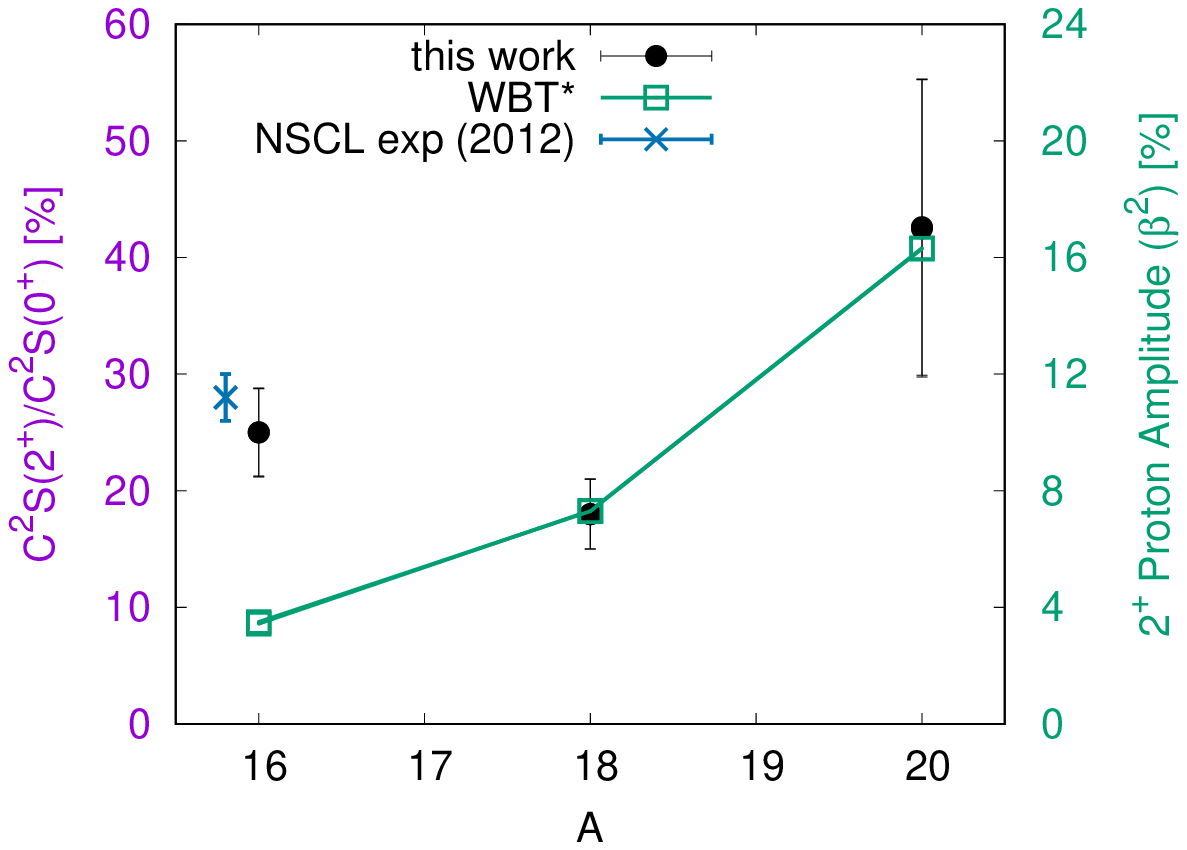}
  \caption{(Color online) Ratio of spectroscopic factors and proton amplitudes for the neutron-rich carbon isotopes ($^{16,18,20}$C) as extracted from this work and how they compare with shell-model calculations using the WBT* interaction \cite{Stanoiu2008,PhysRevC.46.923}. The value for $^{16}$C was also measured in Ref.~\cite{Petri2012}  using a one-proton removal reaction on a $^9$Be target and agrees well with this work.
\label{Fig_ProtonAmplitude}}
\end{figure}

Figure~\ref{Fig_ProtonAmplitude} compares the experimentally derived ratio of spectroscopic factors and proton amplitudes with predictions from the shell model using the WBT* interaction \cite{Stanoiu2008, PhysRevC.46.923}, which reproduces well the excitation spectra of the even-even neutron-rich carbon isotopes. The observed increase in the proton component towards $^{20}$C is well captured within the shell model, which is also successful in explaining the increase in the $E2$ transition probabilities. A similar increase in the proton amplitude is predicted within the seniority scheme of Ref.~\cite{Macchiavelli2014}. 

\begin{figure}[tbp]
\centering
 \includegraphics[angle=0, width=0.7\textwidth, trim=0 0 0 0,clip=true]{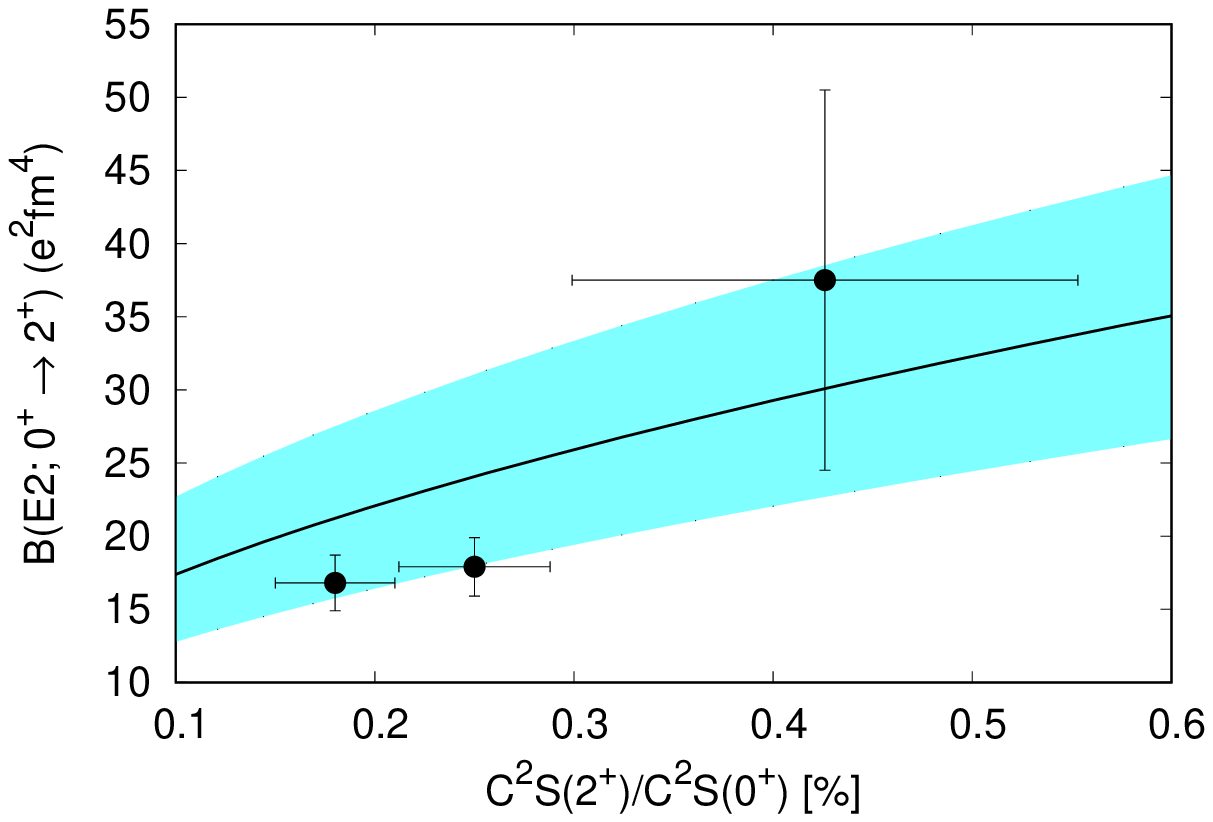}
 \caption{(Color online) Ratio of spectroscopic factors from this work versus transition strengths from Ref.~\cite{PRITYCHENKO20161} for $^{16,18,20}$C. There is a clear correlation that shows that the increased quadrupole strength is due to the enhanced proton contribution to the $2^+_1$ state in $^{20}$C. The shaded area shows the expected limits in the seniority model \cite{Macchiavelli2014} within one standard deviation, when the proton and neutron $E2$ matrix element are fitted to the experimental values of $B(E2)$ and $C^2S(2^+_1) / C^2S(0_1^+)$ for $^{14,16,18,20}$C (see text for details).
\label{Fig_xs_be2}}
\end{figure}

It is interesting to plot the experimental data in the form of Figure~\ref{Fig_xs_be2}, which empirically  shows that the increased quadrupole strength is clearly 
due to the enhanced proton contribution to the $2^+_1$ state in $^{20}$C. For reference, we also include in Figure~\ref{Fig_xs_be2} the results of the seniority model \cite{Macchiavelli2014} if we perform a simultaneous fit of the proton and neutron $E2$ matrix elements to the experimental data of the $E2$ transition probabilities (from Ref.~\cite{PRITYCHENKO20161}) and the ratio of spectroscopic factors for $^{14,16,18,20}$C (from Ref.~\cite{KASCHL1971275} for $^{14}$C and from this work for $^{16,18,20}$C). The black line shows the result of the fit, while the shaded area corresponds to one standard deviation. 

As discussed in the Introduction, the correlation is anticipated on general theoretical arguments that point out a reduction of the $\pi 1p_{1/2} - 1p_{3/2}$ spin-orbit splitting at $\mathrm{Z}=6$ towards the dripline, due to the monopole shifts of the proton effective single-particle levels induced by the successive addition of neutrons in the $sd$ shell.
In first order perturbation theory, the proton amplitude is given by $\beta \sim {V_{\pi\nu}}/{ (E_{2^+_\pi}-E_{2^+_\nu})}$,
with $V_{\pi\nu}$ being the matrix element mixing the unperturbed 2$^+_\pi$ and 2$^+_\nu$ states.
Because the energy denominator,  $ E_{2^+_\pi}-E_{2^+_\nu}$, is dominated by the difference between the proton $1p_{1/2}$ and $1p_{3/2}$ level energies,  $ e_{p1/2}-e_{p3/2}= \Delta E_{so} $,
it is clear that a reduced spin-orbit splitting will help promote proton excitations.

In a more quantitative way, we consider the semi-empirical analysis of the isospin dependence of the  $1p$ spin-orbit splitting in Ref.~\cite{MAIRLE197797}.
According to their results, $\Delta E_{so}$ changes from $\approx 6$~MeV in  $^{14}$C to $\approx 4$~MeV in  $^{20}$C.
These values are in good agreement with the ones predicted from the WBT* interaction ($\Delta E_{so} = 6.54$~MeV at $\rm N=8$ and $4.80$~MeV at $\rm N=14$) and a Woods-Saxon potential with the FSU parametrization \cite{COSMO} ($\Delta E_{so} = 5.7$~MeV at $\rm N=8$ and $4.4$~MeV at $\rm N=14$). It is also interesting to point out that using monopole averages calculated from the Schiffer-True interaction~\cite{STint} and assuming fully mixed $sd$ neutrons,  we estimate a change in the spin-orbit splitting between $^{14}$C and $^{20}$C  $\delta \Delta E_{so} \approx$ -1.75 MeV, consistent with the values above\footnote[1]{ The relevant average proton-neutron matrix element differences are decomposed as: $\langle V_{p_{1/2}-d_{5/2}} \rangle - \langle V_{p_{3/2}-d_{5/2}} \rangle= -0.05 (\rm Central) -0.31 (\rm Tensor) -0.02 (\rm LS)$ MeV and $\langle V_{p_{1/2}-s_{1/2}} \rangle - \langle V_{p_{3/2}-s_{1/2}} \rangle= -0.024 (\rm LS)$ MeV}.

Using the $\Delta E_{so}$ from Ref.~\cite{MAIRLE197797}, we determine an average mixing matrix element 
$V_{\pi\nu} \approx 1.5$ MeV.      
Similarly,  and although excitation energies could be more affected than the amplitudes, we obtain a value of $V_{\pi\nu} \approx 1.2$~MeV from the lowering of the $2^+$ states with respect to their expected unperturbed values, consistent with the estimate above. Using an SDI interaction \cite{Brussaard} with strength parameters $A_{T=0}=A_{T=1}=1.5$~MeV,  scaled to this mass region, the mixing matrix element is calculated \cite{PietVI} to be  $V_{\pi\nu}^{\rm SDI} \approx 1.3$~MeV in line with the empirical results.

Empirically, as shown in Fig. 5, the proton-amplitudes and the $B(E2)$'s correlate well. However, and while the statistics on the cross-section ratios does not allow us to make a firm statement, the data may suggest a reduction (increase) of the proton component in $^{18}$C ($^{16}$C). This apparent decrease (increase) is not anticipated from theoretical arguments and we currently do not have an explanation for this behavior. Perhaps systematic uncertainties,  due to the subtraction of the indirect feeding of the $2^+_1$ and/or the theoretical cross-sections,  are not fully accounted in the quoted errors.

\section{Summary \label{Summary}}

In this work, we have employed $\rm ^AN$(p,2p)$\rm^{A-1}C$ quasi-free scattering reactions using relativistic radioactive beams to study the proton component of the $2^+_1$ state in $^{16,18,20}$C, and look into the evolution of the $\rm Z=6$ spin-orbit splitting towards the neutron dripline. 
Our results show an increase in the proton component, and signal moderate quenching of the $\mathrm{Z}=6$ $1p_{1/2} - 1p_{3/2}$ gap towards the dripline, in contrast to the conclusions of Ref.~\cite{Tran2018}.
The driving mechanism behind the evolution of the $\pi1p_{1/2}$ and $\pi1p_{3/2}$ orbits as function of isospin is the combined effect of the tensor (mainly) and two-body spin-orbit forces acting on the $1p$ protons when neutrons are added in the $d_{5/2}$ and $s_{1/2}$ orbits.
We expect that these results will motivate further theoretical work on the structure of neutron-rich carbon isotopes from both large-scale shell model calculations and ab initio approaches. Experimentally, the study of unbound (mixed-symmetry) 2$^+$ states~\cite{Macchiavelli2014} appears as the next logical step and is underway.


\section*{Acknowledgments}
This work was supported by the 
Royal Society award UF150476,
UK STFC awards ST/M006433/1 and ST/P003885/1,
U.S. Department of Energy, Office of Nuclear Physics, contract No. DE-AC02-05CH11231 (LBNL), the German Federal Ministry of Education and Research (BMBF projects 05P15RDFN1, 05P15WOFNA and 05P19RDFN1), the GSI-TU Darmstadt cooperation agreement, the State of Hesse through the LOEWE center HIC for FAIR, the Helmholtz-Gemeinschaft through the graduate school HGS-HIRe and Young Investigators Grants, and the Swedish Research Council under contract number 621-2011-5324.
This work was also supported by the Spanish research grants (Ministerio Ciencia e Innovaci\'on), FPA2015-64969-P,  FPA2015-69640-C2-1-P,  FPA2017-87568-P, PGC2018-099746-B-C21,  Undidad de Excelencia MarÃ­a de Maeztu under project MDM-2016.0692, and  the Xunta de Galicia research grant GRC  ED431C 2017/54. This work was partly supported by the Deutsche Forschungsgemeinschaft (DFG, German Research Foundation) - Project-ID 279384907 - SFB 1245.
C.A.B. acknowledges support from the U.S. NSF Grant No. 1415656, and U.S. DOE Grant No. DE-FG02-08ER41533.
A.O.M. would like to thank Piet Van Isacker for illuminating discussions regarding the $V_{\pi\nu}$ mixing matrix elements.


\bibliography{carbon}

\end{document}